\documentstyle[prl,aps,twocolumn,floats]{revtex}
\begin{document}
%\draft

\title
{Cosmic Acceleration: A Natural Remedy For Horizon and Flatness Problems}

\author{Abhas Mitra}
\address{Theoretical Physics Division, Bhabha Atomic Research Center,\\
Mumbai-400085, India\\ E-mail: amitra@apsara.barc.ernet.in}

%\date{\today}

\maketitle

\begin{abstract}
The standard big bang cosmology has been greatly successful in explaining
many observational aspects of the real universe. However, two particular
diffficulties faced by it are the so-called ``horizon'' and ``flatness''
problems. By assuming that the recently found cosmic acceleration to be a
genuine and ever present effect, we show that, the resultant modified
Friedman model is free from both the above referred problems, and there is
no need to invoke the additional ``inflationary'' initial phase as a
seperate ingredient. The important cosmological milestones of the standard
big babg cosmology (which predicts cosmic deceleration) like GUT symmetry
breaking epoch or radiation matter decoupling epoch may remain more or
less unchanged in the modified cosmology in terms of redshift values.
However, in terms of cosmic time $t$, all such epochs occur earlier as the
universe becomes older in this case.
\end{abstract}

%PACS:
%\newpage
\vskip 0.5cm
\newpage

\section{Introduction}
Only a few years ago, it was believed that the universe is undergoing a
declerated expansion (DE). Then it was felt that there is no compelling reason
to retain the ``cosmological constant'' term, $\Lambda$, in the Einstein
equations describing the evolution of the universe. In such a case, it
followed that, the behaviour of the universe filled with matter and
radiation, could be broadly described as $S(t)= \alpha t^n$, where $n<1$
and $\alpha$ is a constant of proportionality\cite{1,2}. For instance, for the early
radiation dominated era, one has $S \sim t^{1/2}$ and for the matter
dominated universe, $S\sim t^{2/3}$ (for a closed universe with curvature
parameter $k=0$).

However, over the past few years, studies of luminosity-distance studies
involving type supernova I, have suggested that the universe is undergoing
accelerated expansion (AE). Such AE is possible either due to presence of
a $\Lambda$ term or because of the presence of a quintessence\cite{3}. The
simplest way to broadly reprsent the AE would be to write

\begin{equation}
S(t)= \alpha t^n; \qquad ~ n>1
\end{equation}
In the past era of supposed decelerated expansion, one would encounter the
socalled horizon and flatness puzzle. In the former case, it is difficult
to find an answer asto why the universe is so isotropic and homogeneous if
the universe comprised causally disconnected tiny patches in the very early
era. In the latter case, one has to explain why the quantity $\mid
\Omega -1\mid \sim {\cal O} 1$ in the present epoch unless $\Omega$ was
extremely fine tuned to unity in the early past. Here $\Omega$ is the ratio
of the actual density of the universe ($u$ to the critical density needed
to close the universe$u_c= 8\pi G/3 H^2$, where $H$ is the Hubble constant. To remedy such problems,
in the framework of the supposed (incorrect) decelerated expansion, one
needed to introduce, in a fairly {\it ad hoc} manner, a brief spell of
sudden exponential expansion by a factor of $\sim 10^{30}$ starting from
$t\approx 10^{-35}$ s.   It is this rapid expansion which is supposed to
to bring distant regions within the sphere of causality and simultaneously
flattening the universe\cite{2,3,4,5}. Following this  brief infalationary
period, the universe is supposed to revert back to the original
decelerated expansion mode in a manner which is harly well understood.

However, we shall show below that once we accept the fact the universe is
undergoing AE rather than DE and assume that it was so in the past too,
there is no need to posulate an {\it ad hoc} inflationary phase alteast
for the purpose of solving the horizon or flatness problems.

\section{Removal of Horizon Problem}
The range of the causal horizon is defined as
\begin{equation}
d_H = S(t)\int_0^t {dt'\over S(t')}
\end{equation}
Using Eq. (1) we see that
\begin{equation}
d_H = {S(t)\over k}\int_0^t t'^{-n} dt'={S(t)\over k} {t'^{(1-n)}\over (1-n)}\mid_0^t
\end{equation}
For $n>1$, the above integral blows up and there is no horzon problem.

\section{Flatness Problem}
In the presence of the cosmological term, the relationship between present
density $u$, Hubble constant $H$, and scale size $S$ is\cite{2}
\begin{equation}
H^2+ {K\over S^2} = {8\pi G u\over 3} + {\Lambda c^2\over 3}
\end{equation}
Dividing both sides by $H^2$, we have
\begin{equation}
1+ {K\over H^2 S^2} = \Omega + {\Lambda c^2\over 3 H^2}
\end{equation} 
Or,
\begin{equation}
\Omega -1 = {k\over H^2 S^2}  - {\Lambda c^2\over 3 H^2}
\end{equation}
Here note that
\begin{equation}
{\dot S} = n\alpha  t^{\beta}; \qquad \beta=n-1 >0; \qquad H={{\dot
S}\over S} = n/t
\end{equation}
Using the foregoing relationships, we find from Eq. (6) that for the present
epoch $t=t_1$ and for the past epoch $t=t_2$
\begin{equation}
{\Omega_1 -1\over \Omega_2 -1} = {k \alpha^{-2} t_1^{-2 \beta}  - {\Lambda_1
t_1^2 c^2\over 3 }\over k\alpha^{-2} t_2^{-2 \beta}  - {\Lambda_2 t_2^2
c^2\over 3}}
\end{equation}
As $t_2 \rightarrow 0$, the second term in the denominator of the above
equation $\rightarrow 0$, and we obtain
\begin{equation}
{\Omega_1 -1\over \Omega_2 -1} = \left({t_1\over t_2}\right)^{-2 \beta}  - \left({\Lambda_1
\over \Lambda_2}\right) t_1^2  t_2^{2\beta}
\end{equation}

Again as $t_2 \rightarrow 0$, both the terms on the RHS of the foregoing
equation $\rightarrow 0$ if $\Lambda$ varies less sharply than $t^{-2}$.
Therefore, we find that
\begin{equation}
\Omega_1 -1 = {\cal O} (0)  (\Omega_2 -1)
\end{equation}

So in the present universe, we should not only have $\mid \Omega_1-1\mid
\sim {\cal O} (1)$, but, we should actually have $\Omega_1
=\Omega_{matter} +\Omega_{\Lambda} \rightarrow 1$.
Then there is no flatness problem, and on the other hand, the universe
should be completely flat at any appreciable value of $t$. In fact very
recent observations indeed show that this is precisely the case\cite{6}.

\section{Discussion and Conclusion}
The age of the universe in the present case could be finite. This can be
seen in the following simple way:
\begin{equation}
t=\int {dS\over {\dot S}}
\end {equation}
By using Eq. (7), we see that
\begin{equation}
t = {S^{1/n}\over \alpha}
\end{equation}
In contrast there could be more involved models incorporating AE for which
one may have $t=\infty$ for $S\neq 0$. In other words there could be
singularity free models for which horizon and flatness problems naturally
do not arise\cite{7}. However, we found that in a more orthodox model of AE with a
finite age of the universe, horizon and flatness problems cease to exist
and, thus, there is no need to postulate, in an {\it adhoc} manner the
occurrence of a brief spell of inflation. It can be shown that even if the
horizon and flatness problems disappear with the idea of inflation in the
present epoch, in principle, they may reappear, in future if $k \neq 0$
and the universe
were undergoing DE\cite{8,9}. On the other hand, such problems do no recur for the
present simple model incorporating AE.  And although, the singularity and
a hot early epoch lie
in a finite past, in the present simple case of an orthodox model, note that
the universe begins with a ``whimper'' (${\dot S} =0$ 
infinity) rather than a ``bang''(${\dot S}=\infty$).

We have of course, not suggested any physics which may be driving the
universe in the AE mode. This a completely different problem in its own right
and many authors are addressing to it. We would simply mention here that
it is possible to have practically any form of a time varying $\Lambda$
and consequent desired evolution\cite{10}.  And if we consider the AE occuring due
to a $\Lambda$ term, which is intrinsic to the fabric of spacetime, the
presence of such a term may not significantly alter the local
thermodynamics. Thus the relationship between $S $ or $z$ and temperature
could be unaltered in such a scheme. In other words, as an example, the
epoch of decoupling may still be at $z\approx 1100$ as in the standard
$\Lambda$-free cosmology. However the $S(t)$ or $z$ and $t$ will change.
Consequently the epoch of decoupling will lie at different model dependent
cosmic time.

\end{document}